\documentclass[showpacs]{revtex4} 
\usepackage{epsfig,amsmath,amssymb,graphicx,upgreek,textcomp}
\usepackage{natbib}
\usepackage[english]{babel} %

\begin{document}

\vspace{0mm}
\title{Quantum distribution functions in systems with an arbitrary number of particles  } %
\author{Yu.M. Poluektov}
\email{yuripoluektov@kipt.kharkov.ua (y.poluekt52@gmail.com)} %
\affiliation{National Science Center ``Kharkov Institute of Physics and Technology'', 61108 Kharkov, Ukraine} %
\author{A.A. Soroka} %
\affiliation{National Science Center ``Kharkov Institute of Physics and Technology'', 61108 Kharkov, Ukraine} %

\begin{abstract}
Expressions for the entropy and equations for the quantum
distribution functions in systems of non-interacting fermions and
bosons with an arbitrary, including small, number of particles are
obtained in the paper.
\newline%
{\bf Key words}: %
distribution function, fermions, bosons, entropy, quantum dot,
factorial, gamma function, Bose-Einstein condensation
\end{abstract}
\pacs{%
05.30.--\,d, 05.30.Ch, 05.30.Fk, 05.30.Jp, 64.60.an, 68.65.--\,k }%
\maketitle

\section{Introduction}\vspace{-0mm} 
Currently, much attention is paid to the study of quantum properties
of systems with a small number of particles, such as quantum dots,
other mesoscopic objects and nanostructures. In this regard, the
problem of describing such objects with taking into account their
interaction with the external environment is of current importance.

Statistical description is usually used to study systems with very
large numbers of particles. But statistical methods of description
can also be used in the study of equilibrium states of systems with
a small number of particles and even one particle. When considering
a system within a grand canonical ensemble, it is assumed that it is
a part of a very large system, a thermostat, with which it can
exchange energy and particles. The thermostat itself is
characterized by such statistical quantities as temperature $T$ and
chemical potential $\mu$. Assuming that the subsystem under
consideration is in thermodynamic equilibrium with the thermostat,
the subsystem itself is characterized by the same values, even if it
consists of a small number of particles. For example, we can
consider the thermodynamics of an individual quantum oscillator [1].
In the case when an exchange of particles with a thermostat is
possible, the time-averaged number of particles of a small subsystem
may be not an integer and, in particular, even less than unity.

In statistical physics, the entropy and distribution functions of
particles over quantum states are calculated under the assumption
that the number of particles is very large. This consideration for
fermions leads to the Fermi-Dirac distribution (FD), and for bosons
-- to the Bose-Einstein distribution (BE) [1].

In this work, the entropy and distribution functions of
non-interacting particles are calculated in the case when no
restrictions are imposed on their number in a system being in
thermodynamic equilibrium with the environment. In particular, the
number of particles can be small, and not an integer and even less
than unity. Equations determining the distribution functions of
fermions and bosons are obtained and their differences from the
standard FD and BE distributions are analyzed. A feature of the
obtained exact distribution functions, in comparison with the
distributions found in the limit of a large number of particles, is
the presence of energy boundaries, beyond which the average number
of particles at all levels turns to zero or unity.

\section{Entropy and distribution function of fermions }\vspace{-0mm} %
Let us consider a quantum system of fermions whose energy levels
$\varepsilon_j$ have the multiplicity of degeneracy $z_j$. If at
each level $j$ there are $N_j\leq z_j$ particles, then the
statistical weight of such a state in the case of FD statistics is
given by the well-known formula [1]
\begin{equation} \label{01}
\begin{array}{l}
\displaystyle{%
   \Delta\Gamma_j=\frac{z_j!}{N_j!\big(z_j-N_j\big)!}\,.  %
}
\end{array}
\end{equation}
The entropy is defined as the logarithm of the total statistical
weight by the relation
\begin{equation} \label{02}
\begin{array}{l}
\displaystyle{%
   S=\ln\Delta\Gamma =\sum_j\ln\Delta\Gamma_j= %
   \sum_j\!\Big[\ln z_j!-\ln N_j!-\ln\!\big(z_j-N_j\big)!\Big]. %
}
\end{array}
\end{equation}
To calculate all factorials under assumption $N\gg 1$ and $z\gg 1$,
the Stirling's formula [2] is usually used in the form
\begin{equation} \label{03}
\begin{array}{l}
\displaystyle{%
    \ln N! \approx N \ln\!\Big(\frac{N}{e}\Big).  %
}
\end{array}
\end{equation}
When studying systems with small $N$, the accuracy of this formula
is insufficient. So, for example, with $N=16$ its accuracy is 7.5\%.
For $N=1,\,2$ there are negative numbers on the right in (3). For
small $N$, one can use a more accurate formula
\begin{equation} \label{04}
\begin{array}{l}
\displaystyle{%
    \ln N! \approx N \ln\!\Big(\frac{N}{e}\Big)+\ln\sqrt{2\pi N}.  %
}
\end{array}
\end{equation}
For $N=16$ its accuracy is already 0.017\%, and for $\ln2!\approx 0.693$ %
this formula gives a value of 0.652. Taking into account the more
accurate formula (4), for the entropy $S=\sum_j S_j$ we obtain the expression %
\begin{equation} \label{05}
\begin{array}{l}
\displaystyle{%
   S_j=-z_j\big[n_j\ln n_j + (1-n_j)\ln(1-n_j)\big] -\frac{1}{2}\ln\!\big[2\pi z_jn_j(1-n_j)\big]. %
}
\end{array}
\end{equation}
Here $n_j=N_j\big/z_j$ is the average number of particles at level
$j$ or the population of the level. This formula differs from the
usual formula for the entropy of a Fermi gas [1] by the last term.
Taking into account that the total number of particles $N$ and the
total energy $E$ are determined by the formulas
\begin{equation} \label{06}
\begin{array}{l}
\displaystyle{%
   N=\sum_jN_j=\sum_jn_jz_j,  %
}
\end{array}
\end{equation} \vspace{-5mm}
\begin{equation} \label{07}
\begin{array}{l}
\displaystyle{%
   E=\sum_j\varepsilon_jN_j=\sum_j\varepsilon_jn_jz_j,  %
}
\end{array}
\end{equation}
the average number of particles in each state is found from the condition %
\begin{equation} \label{08}
\begin{array}{l}
\displaystyle{%
   \frac{\partial}{\partial n_j}\big(S-\alpha N-\beta E\big)=0,  %
}
\end{array}
\end{equation}
where $\alpha,\beta$ are the Lagrange multipliers. From here we find
the equation that determines the average number of particles in a state $j$: %
\begin{equation} \label{09}
\begin{array}{l}
\displaystyle{%
   \ln\!\frac{1-n_j}{n_j}+\frac{1}{2z_j}\Big(\frac{1}{1-n_j}-\frac{1}{n_j}\Big) = \alpha+\beta\varepsilon_j\equiv\theta_j.  %
}
\end{array}
\end{equation}
Neglecting the second term on the left side, we obtain the usual FD distribution %
\begin{equation} \label{10}
\begin{array}{l}
\displaystyle{%
   n_j^{(0)}=\frac{1}{ \vphantom{^{1^1}} e^{\alpha+\beta\varepsilon_j}+1 }. %
}
\end{array}
\end{equation}
From comparison with thermodynamic relations it follows that
$\alpha=-\mu/T$, $\beta=1/T$, $T$ -- temperature, $\mu$ -- chemical
potential, so that $\theta_j=\big(\varepsilon_j-\mu\big)\big/T$. In
the absence of a magnetic field, for particles with spin 1/2 the
smallest multiplicity of degeneracy only in the spin projection is
equal to two. With $z_j\gg 1$, the second term on the left in (9)
can be taken into account as a small correction, so that in this
approximation the distribution function will take the form
\begin{equation} \label{11}
\begin{array}{l}
\displaystyle{%
   n_j=n_j^{(0)}-\frac{\big(1-2n_j^{(0)}\big)}{2z_j}. %
}
\end{array}
\end{equation}
The domain of variation of the parameter $\theta_j$ is determined by
the condition $0\le n(\theta_j)\le 1$.

For an arbitrary, including small and non-integer number of
particles $N$, the factorial should be determined through the gamma
function $\Gamma(x)$:
\begin{equation} \label{12}
\begin{array}{l}
\displaystyle{%
   N! = \Gamma(N+1). %
}
\end{array}
\end{equation}
In this case the statistical weight (1) is also expressed through
the gamma function:
\begin{equation} \label{13}
\begin{array}{l}
\displaystyle{%
   \Delta\Gamma_j = \frac{\Gamma(z_j+1)}{\Gamma(N_j+1)\Gamma(z_j-N_j+1)}. %
}
\end{array}
\end{equation}
Some formulas for the gamma function and related to it formulas are
given in the Appendix. With allowance for (13), for the
nonequilibrium entropy $S=\sum_jS_j$ there follows the formula
\begin{equation} \label{14}
\begin{array}{l}
\displaystyle{%
   S_j = -\ln\Gamma(z_jn_j+1)-\ln\Gamma\big[z_j(1-n_j)+1\big] + \ln\Gamma(z_j+1). %
}
\end{array}
\end{equation}
It is obvious that the contribution to the total entropy comes only
from partially occupied levels, for which $0<n_j<1$. In this case,
when $n_j\neq 0,1,$\, from the condition (8) we find the equation
that determines the average number of particles in each state
\begin{equation} \label{15}
\begin{array}{l}
\displaystyle{%
   \psi\big[z_j(1-n_j)+1\big] -\psi\big(z_jn_j+1\big) \equiv \theta_j(n_j)=\frac{(\varepsilon_j-\mu)}{T},  %
}
\end{array}
\end{equation}
where $\psi(z)$  is the logarithmic derivative of the gamma function
(the psi function) (A4). If we use the asymptotic formulas (A3) and
(A6) given in the Appendix, then formula (14) will turn into (5),
and formula (15) into (9). Using formula (A11), equation (15) can be
written in the form
\begin{equation} \label{16}
\begin{array}{l}
\displaystyle{%
   z_j(1-2\,n_j)\sum_{k=1}^\infty\frac{1}{\big[k+z_j(1-n_j)\big]\!\big[k+z_jn_j\big]} = \frac{(\varepsilon_j-\mu)}{T}.  %
}
\end{array}
\end{equation}
Note that here the series converges rather slowly and the rate of
its convergence decreases with increasing $z_j$, so that for
calculations it is more convenient to use formula (15).

The form of distribution functions for a system of Fermi particles
at $z=2$, $z=10$ and arbitrary $j$ is shown in Fig.\,1. The
dependence $n(\theta)$, obtained from equation (9) with %
$\theta_j=\ln\!\big[(1-n_j)/n_j\big]+(1/2z_j)\big[(1-n_j)^{-1}-n_j^{-1}\big]$ %
(curves {\it 2} in Fig.\,1), turns out to be multiple-valued and
leads to a significant discrepancy with the calculation performed
using the exact formula (15) (curves {\it 1} in Fig.\,1), so that
equation (9) turns out to be inapplicable for calculating average
occupation numbers. In the standard FD distribution (10) (curve {\it
4} in Fig.\,1), for an arbitrary value of the parameter
$-\infty<\theta<\infty$ the distribution function does not turn
exactly to zero or unity. At $\theta\rightarrow\infty$ the
distribution function exponentially tends to zero $n(\theta)\sim
e^{-\theta}$, and at $\theta\rightarrow-\infty$ it tends to unity
$n(\theta)\sim 1-e^\theta$.
\vspace{-1mm} %
\begin{figure}[h!]
\vspace{-0mm}  \hspace{0mm}
\includegraphics[width = 15cm]{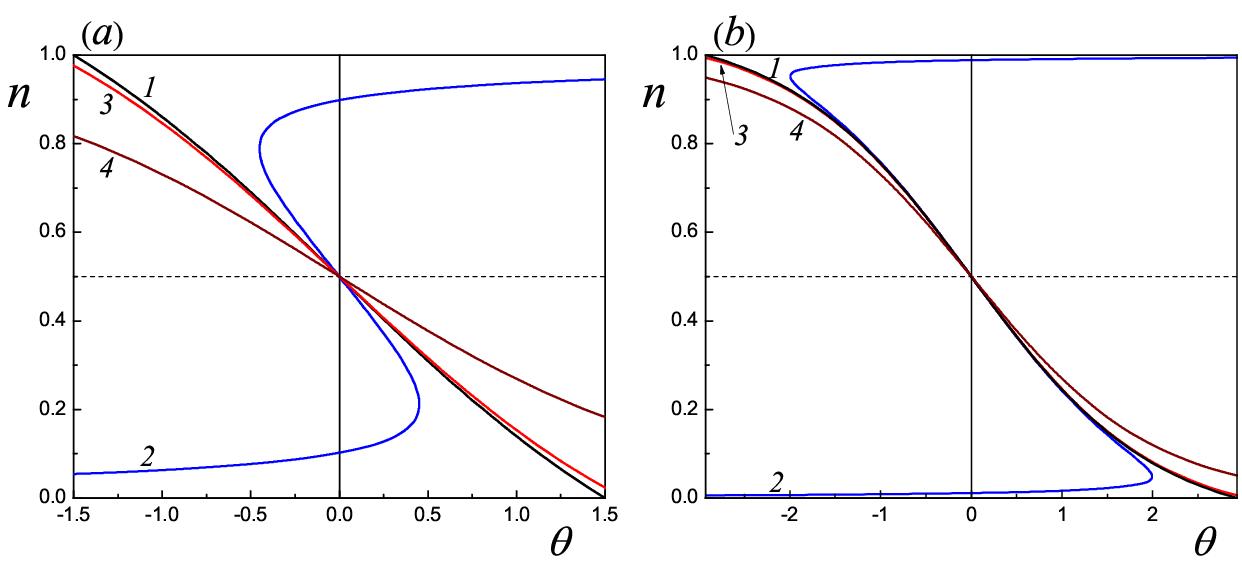} 
\vspace{-3mm} %
\caption{\label{fig01} 
Distribution function of Fermi particles $n(\theta)$ over states in
various approximations with multiplicities of level degeneracy: %
(\!{\it a}) $z=2$, (\!{\it b}) $z=10$.
{\it 1} -- distribution function (DF), calculated using the exact formula (15); %
{\it 2} -- DF, calculated using approximate equation (9); %
{\it 3} -- DF, calculated using formula (11); %
{\it 4} -- conventional Fermi-Dirac DF (10). %
}%
\end{figure}

A feature of the exact distribution function defined by equations
(15),\,(16) is the limited range of values of the parameter
$\theta_j$, in which the function is different from zero or unity. %
In this case $\theta_{j\,{\rm min}}<\theta_j<\theta_{j\,{\rm max}}$,
where
\begin{equation} \label{17}
\begin{array}{l}
\displaystyle{%
   \theta_{j\,{\rm max}}=-\theta_{j\,{\rm min}}=\psi\big(z_j+1\big)-\psi\big(1\big) = \sum_{k=1}^{z_j}k^{-1}.  %
}
\end{array}
\end{equation}
At $\theta_j\ge\theta_{j\,{\rm max}}$ the average number of
particles at level $j$ becomes zero $n_j=0$, and at
$\theta_j\le\theta_{j\,{\rm min}}$ it is equal to unity $n_j=1$.
Thus, for given values of $T$ and $\mu$, the population of level $j$
is different from zero and unity when there is fulfilled the inequality %
\begin{equation} \label{18}
\begin{array}{l}
\displaystyle{%
   -\theta_{j\,{\rm max}}< \frac{\varepsilon_j-\mu}{T} <\theta_{j\,{\rm max}}.  %
}
\end{array}
\end{equation}
All the other levels remain either empty or completely occupied, %
so that there is only a finite number of partially occupied levels
and their number increases with increasing temperature.

The approximate expression for the distribution function (11)
(curves {\it 3} in Fig.\,1) following from formula (9) gives a good
approximation to the exact dependence (curves {\it 1} in Fig.\,1).
However, at points where the exact distribution function becomes
zero and unity, the approximate function (11) is different from
these values and exists for all values of the parameter $\theta$. %
The difference between the exact distribution (15) (curves {\it 1}
in Fig.\,1) and the usual FD distribution (10) (curves {\it 4} in
Fig.\,1) is more significant the larger the absolute value of the
parameter $\theta$ and the smaller the degeneracy factor $z$.

Equation (15) and approximate formula (11) determine the average
number of particles in a state $j$ as a function of temperature and
chemical potential $n_j=n_j(T,\mu)$. A substitution of these
functions into (6),\,(7),\,(14) gives equilibrium values of the
entropy $S=S(T,\mu)$, the energy $E=E(T,\mu)$ and the number of
particles $N=N(T,\mu)$ as functions of temperature and chemical
potential. These quantities are natural variables for the large
thermodynamic potential, which can be defined by the usual expression %
\begin{equation} \label{19}
\begin{array}{l}
\displaystyle{%
   \Omega(T,\mu)=E(T,\mu)-TS(T,\mu)-\mu N(T,\mu),  %
}
\end{array}
\end{equation}
so that at a constant volume there holds the well-known identity
$d\Omega=-SdT-Nd\mu$. For a fixed total number of particles
equations (15) are not independent, since the populations of the
levels are linked by the relation (6). If the total number of
particles is such that they can completely occupy the lower $j$
levels, and the level $j+1$ turns out to be occupied partially, so that %
$N=\sum_{k=1}^jz_k+N_{j+1}$ and $0<N_{j+1}<z_{j+1}$, then at
$T\rightarrow 0$ the chemical potential takes the value $\mu=\varepsilon_{j+1}$. %
Near zero temperature $\mu=\varepsilon_{j+1}-T\theta_{j+1}\big(N_{j+1}/z_{j+1}\big)$. %
The entropy at zero temperature turns to zero only in the case when
all levels are completely occupied or empty. In the presence of an
unoccupied level the entropy at $T= 0$ is different from zero. %
Thus, the third law of thermodynamics is always satisfied in the
Nernst formulation, according to which all processes at zero
temperature occur at a constant entropy. And in the Planck
formulation, which requires turning of the entropy to zero, the
third law is satisfied only in the case of completely occupied levels. %

\section{Entropy and distribution function of bosons }\vspace{-0mm} %
If at each level of a boson system with the multiplicity of
degeneracy $z_j$ there are $N_j$ particles, then the statistical
weight of such a state in the BE statistics [1]
\begin{equation} \label{20}
\begin{array}{l}
\displaystyle{%
   \Delta\Gamma_j = \frac{(z_j+N_j-1)!}{N_j!\,(z_j-1)!}. %
}
\end{array}
\end{equation}
The entropy is defined by the relation
\begin{equation} \label{21}
\begin{array}{l}
\displaystyle{%
   S=\ln\Delta\Gamma =\sum_jS_j=\sum_j\ln\Delta\Gamma_j= %
   \sum_j\!\Big[\ln\!\big(z_j+N_j-1\big)!-\ln N_j!-\ln\!\big(z_j-1\big)!\Big]. %
}
\end{array}
\end{equation}
It should be noted that if the level is not degenerate $z_j=1$ or
not occupied $N_j=0$, then, as in the considered above case of
Fermi-Dirac statistics (1) $\Delta\Gamma_j=1$, and it does not
contribute to the total entropy. Using the Stirling's formula (4) we
have
\begin{equation} \label{22}
\begin{array}{l}
\displaystyle{%
   S_j=\big(z_j+z_jn_j-1\big)\ln\!\big(z_j+z_jn_j-1\big) - z_jn_j\ln\!\big(z_jn_j\big) - \big(z_j-1\big)\ln\!\big(z_j-1\big)+  %
}\vspace{2mm}\\ %
\displaystyle{\hspace{5mm}%
  +\,\frac{1}{2}\ln\!\big[2\pi(z_j+z_jn_j-1\big)\big] - \frac{1}{2}\ln\!\big(2\pi z_jn_j\big) - \frac{1}{2}\ln\!\big[2\pi(z_j-1\big)\big]. %
}%
\end{array}
\end{equation}
Then, from condition (8) it follows the equation for the
distribution function over states:
\begin{equation} \label{23}
\begin{array}{l}
\displaystyle{%
   \ln\!\frac{(z_j+z_jn_j-1)}{z_jn_j}+\frac{1}{2z_j}\bigg(\frac{z_j}{z_j+z_jn_j-1}-\frac{1}{n_j}\bigg) = \alpha+\beta\varepsilon_j.  %
}
\end{array}
\end{equation}
For $z_j\rightarrow\infty$, from (23) we obtain the usual BE distribution %
\begin{equation} \label{24}
\begin{array}{l}
\displaystyle{%
   n_j^{(0)}=\frac{1}{ \vphantom{^{1^1}} e^{\alpha+\beta\varepsilon_j}-1 }. %
}
\end{array}
\end{equation}
Taking into account the correction of order $1/z_j$, we have
\begin{equation} \label{25}
\begin{array}{l}
\displaystyle{%
   n_j=n_j^{(0)}-\frac{\big(1+2n_j^{(0)}\big)}{2z_j}. %
}
\end{array}
\end{equation}

When determining factorials through the gamma function, the
statistical weight (20) will be written in the form
\begin{equation} \label{26}
\begin{array}{l}
\displaystyle{%
   \Delta\Gamma_j = \frac{\Gamma(z_j+N_j)}{\Gamma(N_j+1)\,\Gamma(z_j)}. %
}
\end{array}
\end{equation}
This implies the formula for nonequilibrium entropy $S=\sum_jS_j$\,: %
\begin{equation} \label{27}
\begin{array}{l}
\displaystyle{%
   S_j=\ln\Gamma\big(z_jn_j+z_j\big) - \ln\Gamma\big(z_jn_j+1\big) - \ln\Gamma\big(z_j\big). %
}
\end{array}
\end{equation}
As was noted, unoccupied levels $n_j=0$ do not give a contribution to the total entropy. %
For $n_j\neq 0$ from condition (8) we find the equation for the
average number of particles in each state
\begin{equation} \label{28}
\begin{array}{l}
\displaystyle{%
   \psi\big(z_jn_j+z_j\big) - \psi\big(z_jn_j+1\big)\equiv\theta_j(n_j)=\frac{(\varepsilon_j-\mu)}{T}. %
}
\end{array}
\end{equation}
If one uses asymptotic formulas (A3) and (A6), formula (27) will
turn into (22), and formula (28) into (23). By using formula (A8),
equation (28) can be represented as
\begin{equation} \label{29}
\begin{array}{l}
\displaystyle{%
   \sum_{k=1}^{z_j-1}\frac{1}{z_jn_j+z_j-k} = \frac{(\varepsilon_j-\mu)}{T}.  %
}
\end{array}
\end{equation}
The form of distribution functions for a system of Bose particles at
$z=2$,$z=10$ and arbitrary $j$ is shown in Fig.\,2. \nolinebreak[4]

In the standard BE distribution (24) the parameter $\theta$ can take
arbitrary positive values $0<\theta<\infty$. At
$\theta\rightarrow\infty$ the distribution function exponentially
tends to zero $n(\theta)\sim e^{-\theta}$, and at $\theta\rightarrow 0$ %
it increases according to the law $n(\theta)\sim 1/\theta$. The
dependence $n(\theta)$ obtained from equation (23) with %
$\theta_j=\ln\!\big[(z_j+z_jn_j-1)/z_jn_j\big]+(1/2z_j)\big[z_j/(z_j+z_jn_j-1)-1/n_j\big]$ %
(curves {\it 2} in Fig.\,2) turns out to be multiple-valued and
leads to a significant discrepancy with the calculation performed
using the exact formula (28) (curves {\it 1} in Fig.\,2). Therefore,
equation (23) is not applicable for calculating average occupation
numbers. However, the approximate formula for the distribution
function (25) following from (23) gives a good approximation, almost
coinciding with the exact dependence (curves {\it 1} in Fig.\,2). %
The essential difference consists in that at some boundary value
$\theta=\theta_{{\rm max}}$ the exact function (28) turns to zero,
while the approximate function (25) remains finite, although exponentially small. %
\vspace{0mm} %
\begin{figure}[h!]
\vspace{-1mm}  \hspace{0mm}
\includegraphics[width = 15cm]{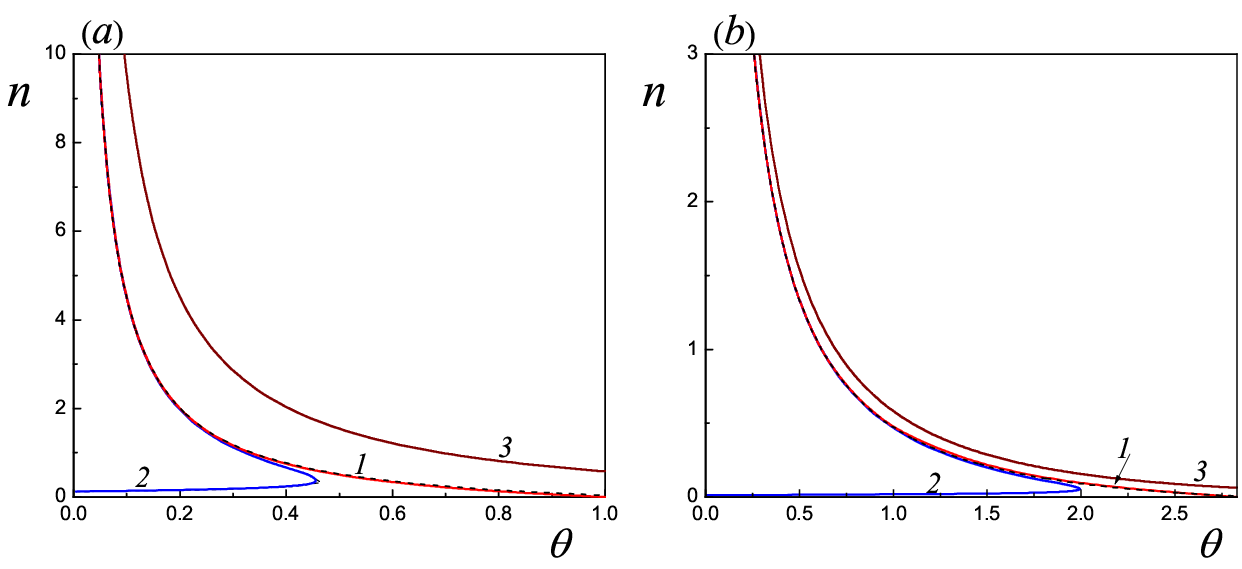} 
\vspace{-3mm} %
\caption{\label{fig01} 
Distribution function of Bose particles $n(\theta)$ over states in
various approximations with multiplicities of level degeneracy: %
(\!{\it a}) $z=2$, (\!{\it b}) $z=10$.
{\it 1} -- distribution function (DF), calculated using the exact formula (28); %
{\it 2} -- DF, calculated using approximate equation (23); %
{\it 3} -- conventional Bose-Einstein DF (24). %
The calculation using the approximate formula (25) gives a
dependence that practically coincides with curve {\it 1} (dotted line). %
}%
\end{figure}

Thus, for the exact DF (28) with values $n_j\neq 0$ the parameter
$\theta_j$ changes in the finite region $0<\theta_j<\theta_{j\,{\rm max}}$, where %
\begin{equation} \label{30}
\begin{array}{l}
\displaystyle{%
   \theta_{j\,{\rm max}}= \psi\big(z_j\big)-\psi\big(1\big)=\sum_{k=1}^{z_j-1}k^{-1}  %
}
\end{array}
\end{equation}
and $n_j\big(\theta_{j\,{\rm max}}\big)=0$. At $\theta_j\rightarrow 0$ %
the exact DF increases according to the law $n_j(\theta_j)\sim (z_j-1)\big/\!z_j\theta_j$. %
Thus, at given $T$ and $\mu$ the population of level $j$ is
different from zero when there holds the inequality
\begin{equation} \label{31}
\begin{array}{l}
\displaystyle{%
   0 < \frac{\varepsilon_j-\mu}{T}<\theta_{j\,{\rm max}}.  %
}
\end{array}
\end{equation}
All the other levels remain empty, so that there is only a finite
number of partially occupied levels and their number increases with
increasing temperature.

At zero temperature and a fixed number of Bose particles only the
lower level is occupied and $\mu=\varepsilon_1$. %
The entropy $S_1=\ln\Gamma(N+z_1)-\ln\Gamma(N+1)-\ln\Gamma(z_1)$
remains nonzero in this case, so that the third law of
thermodynamics is satisfied only in the Nernst formulation. %
With a slight increase in temperature, the lower level continues to
remain occupied with other levels being empty in a certain
temperature range, and the chemical potential changes linearly with
temperature $\mu=\varepsilon_1-T\theta_1(N/z_1)$. %
At the temperature $T_1$, determined by the condition
\begin{equation} \label{32}
\begin{array}{l}
\displaystyle{%
   \frac{\varepsilon_2-\varepsilon_1}{T_1}=\theta_2(0)-\theta_1(N/z_1),   %
}
\end{array}
\end{equation}
there begins filling of the second level, and the number of
particles at the lower level decreases. At further increase of
temperature there begins filling of higher levels. %
At a certain temperature $T_B$ the number of particles at the ground
level will turn to zero. This temperature is determined by the
equations for
partially occupied levels %
\begin{equation} \label{33}
\begin{array}{l}
\displaystyle{%
   \theta_2\big(n_2(T_B)\big)-\theta_{1{\rm max}}= \frac{\varepsilon_2-\varepsilon_1}{T_B}, \quad %
   \theta_3\big(n_3(T_B)\big)-\theta_{1{\rm max}}= \frac{\varepsilon_3-\varepsilon_1}{T_B}, \,\,\ldots\,\,, %
}
\end{array}
\end{equation}
where $\theta_{1{\rm max}}=\psi(z_1)-\psi(1)$, provided that %
$z_2n_2(T_B)+z_3n_3(T_B)+\ldots=N$, \,$\mu_B=\varepsilon_1-T_B\theta_{1{\rm max}}$. %
If one goes down in temperature, then $T_B$ corresponds to the
temperature at which the filling of the lower level begins, and
therefore it can be considered as an analogue of the temperature of
Bose-Einstein condensation in macroscopic systems [1].

\section{Summary and conclusions}\vspace{-2mm} %
In connection with the intensive study of quantum systems of small
sizes, the problem of theoretical description of such objects with
taking into account their interaction with the environment is
becoming increasingly actual. Methods of statistical mechanics can
be used to describe also such systems with a small number of
particles that are in thermodynamic equilibrium with a thermostat.
In this work, the equations are obtained for the distribution
functions of fermions and bosons over quantum states for systems
with an arbitrary, including a small, number of particles. In the
limiting case, when the number of particles and the multiplicity of
degeneracy of levels in the system are large, these distributions
transform into the well-known Fermi-Dirac and Bose-Einstein
distributions. For systems with a small number of particles and at
low temperatures, the average number of particles at a given level
can differ significantly from the values predicted by the FD and BE
distributions. It is of interest to experimentally test the
applicability of the obtained distributions for structures with a
small number of particles.

\appendix

\section{Some properties of the gamma function and \newline\mbox{\quad} its logarithmic derivative (the psi-function)} %
\vspace{-3mm}%
The definition of the gamma function:
\begin{equation} \label{A01}
\begin{array}{l}
\displaystyle{%
   \Gamma(x)=\int_0^\infty\!e^{-t}t^{x-1}dt = \int_0^1\!\Big(\ln\!\frac{1}{t}\Big)^{\!x-1}dt. %
}
\end{array}
\end{equation}
For the logarithm of the gamma function, there holds the integral representation %
\begin{equation} \label{A02}
\begin{array}{l}
\displaystyle{%
   \ln\Gamma(x)=\int_0^\infty\!\bigg[(x-1)e^{-t}-\frac{e^{-t}-e^{-xt}}{1-e^{-t}}\bigg]\frac{dt}{t} %
}
\end{array}
\end{equation}
and the asymptotics at $x\rightarrow\infty$
\begin{equation} \label{A03}
\begin{array}{l}
\displaystyle{%
   \ln\Gamma(x)\sim\bigg(x-\frac{1}{2}\bigg)\ln x - x +\frac{1}{2}\ln(2\pi). %
}
\end{array}
\end{equation}
The logarithmic derivative of the gamma function (the psi function)
is defined by the formula
\begin{equation} \label{A04}
\begin{array}{l}
\displaystyle{%
   \psi(x)\equiv\frac{d}{dx}\ln\Gamma(x)=\frac{1}{\Gamma(x)}\frac{d\Gamma(x)}{dx}. %
}
\end{array}
\end{equation}
The integral representation is valid for it
\begin{equation} \label{A05}
\begin{array}{l}
\displaystyle{%
   \psi(x)=\int_0^\infty\!\bigg[\frac{e^{-t}}{t}-\frac{e^{-xt}}{1-e^{-t}}\bigg]dt %
}
\end{array}
\end{equation}
and the asymptotics at $x\rightarrow\infty$
\begin{equation} \label{A06}
\begin{array}{l}
\displaystyle{%
   \psi(x)\sim\ln x - \frac{1}{2x}. %
}
\end{array}
\end{equation}
There are useful formulas:
\begin{equation} \label{A07}
\begin{array}{l}
\displaystyle{%
   \psi(x+1)=\psi(x)+\frac{1}{x},  %
}
\end{array}
\end{equation}
\vspace{-5mm}%
\begin{equation} \label{A08}
\begin{array}{l}
\displaystyle{%
   \psi(x+n)=\sum_{k=1}^{n-1}\frac{1}{(n-k)+x} + \psi(x+1), \quad (n\ge 2),  %
}
\end{array}
\end{equation}
\vspace{-3mm}%
\begin{equation} \label{A09}
\begin{array}{l}
\displaystyle{%
   \psi(1-x)=\psi(x)+\pi\,{\rm ctg}(\pi x),   %
}
\end{array}
\end{equation}
\vspace{-5mm}%
\begin{equation} \label{A10}
\begin{array}{l}
\displaystyle{%
   \psi(1)=-\gamma, \qquad \psi(n)=-\gamma+\sum_{k=1}^{n-1}k^{-1}, \quad (n\ge 2),    %
}
\end{array}
\end{equation}
\vspace{-3mm}%
\begin{equation} \label{A11}
\begin{array}{l}
\displaystyle{%
   \psi(1+x)=-\gamma + \sum_{k=1}^\infty\frac{x}{k(k+x)},     %
}
\end{array}
\end{equation}
where $\gamma=0.5772$ is Euler's constant.


\end{document}